\RequirePackage{fix-cm}
\documentclass[twocolumn]{svjour3}          
\smartqed  
\usepackage{graphicx}
\usepackage{mathptmx}      
\usepackage{color}
\usepackage{amsmath}
\usepackage{makecell}
\usepackage{ulem}
\usepackage{nicefrac}       
\begin{document}

\title{ Modulation of Neuronal Firing Modes by Electric Fields in a Thermosensitive FitzHugh-Nagumo Model
}


\author{ Ediline L. Fouelifack Nguessap       \and
         Antonio C. Roque \and  Fernando F. Ferreira
}


\institute{E. L. F. Nguessap, A.C. Roque, F.F. Ferreira$^*$ \at
              Department of Physics,FFCLRP-University ~of~ Sao Paulo,\\~14040900,~ Ribeirao ~Preto-SP, Brazil \\
              *Corresponding ~author: \email{ferfff@usp.br}           
   } 

\date{Received: date / Accepted: date}

\maketitle

\begin{abstract}
We introduce a thermosensitive FitzHugh-Nagumo neuron model extended with a third dynamic variable representing an external electric field. This unified framework enables a biophysically grounded analysis of how combined thermal and electrical stimuli modulate neuronal excitability. The model incorporates temperature dependence, ion charge density, cell radius, and voltage-driven stimulation, capturing their joint effects on membrane polarization and firing dynamics. Using bifurcation analysis, Lyapunov exponents, and interspike interval variability, we identify transitions between spiking, bursting, and chaotic regimes. We further show how periodic electric fields tune these dynamics as a function of stimulus amplitude, frequency, and cellular geometry.  Our results provide mechanistic insights into neuronal excitability and suggest avenues for controlling neural activity via hybrid thermal-electrical modulation, with potential applications in neuromodulation therapies and bioelectronics.

\end{abstract}

\section{Introduction}
\label{intro}
Understanding the intricate mechanisms that govern the electrical activity of individual neurons remains a fundamental challenge in neuroscience. The behavior of a single neuron is shaped by the delicate interplay of ion channels, membrane potentials, and external stimuli. This fine-tuned process is driven by ionic currents, particularly the movement of ions such as calcium, potassium, and sodium, which generate the action potentials necessary for signal propagation within the nervous system, a cornerstone of neural communication, as described by foundational work \cite{hahn2001bistability,gu2014potassium}. Several mathematical models, including the well-known Hodgkin-Huxley, FitzHugh-Nagumo (FHN), Morris-Lecar, Hindmarsh-Rose models, have been developed to capture these electrophysiological phenomena, offering a framework to explore how neurons encode and transmit information \cite{oja1982simplified,rall1962electrophysiology,nagumo1972response,achard2006complex,tsodyks1996population,nowotny2007dynamical,hindmarsh1982model}.\\
\quad The behavior of a neuron is far from static; it is highly sensitive to various factors, including its internal dynamics and external influences. External perturbations, including electric fields, light, and temperature, can have profound effects on neural activity, significantly altering the dynamics of a single neuron. Weak electric fields, in particular, influence the movement of ions across neuronal membranes, which can modulate membrane potentials and thereby impacting the generation of action potentials \cite{bikson2004effects,radman2007spike}. For example, Ma et al. \cite{ma2019model,hou2021estimate} introduced a neuron model that accounts for the effects of an external electric field, providing a deeper understanding of how neuronal activity responds to such stimuli. Beyond electric fields, neurons are also sensitive to other environmental factors. The integration of a magnetic flux, as proposed by Lv and Ma \cite{lv2016multiple}, highlights how ions moving through the membrane potential can induce a magnetic field, which subsequently influences the neuron's action potentials.These models have opened the door to exploring more complex phenomena like electromagnetic induction and polarization during ion charge displacement. Thus, many electronic components have been incorporated into the neuronal circuit to produce diverse patterns.  For example, memeristors have been introduced into the Hindmarsh-rose model\cite{bao2019hidden}, and they were able to investigate how this magnetic field affects the electrical activity of neurons, expanding our understanding of neuron behavior under different external influences. These components have also been introduced in the Morris-Lecar model\cite{fan2023firing,bao2023three,li2023coexisting}, in the FitzHugh-Nagumo model\cite{chen2023hidden,chen2023memristor}. So, the incorporation of additional variables, such as electric or magnetic fields, into these neuron models has provided deeper insights into how neurons respond to complex environments, bridging the gap between abstract theoretical models and biological reality.
Beyond electric and magnetic fields, light and temperature have also been shown to significantly modulate neural behavior. Liu et al. \cite{liu2020new} proposed a photosensitive neuron model by introducing a photocell into the FitzHugh-Nagumo framework. This allowed them to simulate the effects of light signals on neuronal activity, demonstrating how time-varying light sources can modulate firing patterns and generate complex dynamics. Thermosensitive neurons, which play an essential role in sensory perception, have been modeled using thermistors integrated into neural circuits. Xu et al. \cite{xu2020dynamics} developed a model in which temperature changes control neuron firing patterns. Depending on the parameters, the model presents various behaviors that range from regular bursting to chaotic firing. This model underscores the sensitivity of neural circuits to external temperature fluctuations and how this can lead to different neuronal responses.  However, these environmental stimuli have often been studied in isolation, limiting our understanding of their potential interactions.\\
Recent advances in neurophysiology have shown that neurons exhibit non-linear dynamics, producing complex responses even under weak external stimuli. The introduction of external perturbations, whether through electrical fields, light, or temperature, adds another layer of complexity, influencing neuronal behaviors such as spiking, bursting, or chaotic activity. These mechanisms are essential not only for understanding the dynamic of a single neuron but also for investigating how larger neural networks function \cite{zandi2020different,xiao2016spatiotemporal,wang2016spatiotemporal,qin2018field,simo2021chimera,hussain2021synchronization,zhu2021effects}.\\
More recent studies have revealed increasingly rich behaviors in thermosensitive neuron models. Hussain et al.\cite{hussain2021chimera} analyzed chimera states in a network of thermosensitive FHN neurons, highlighting the influence of thermal sensitivity on spatiotemporal synchronization. Xing et al. \cite{xing2020bifurcations} investigated bifurcations and excitability in a temperature-sensitive Morris–Lecar neuron, while Guo et al. \cite{guo2022desynchronization} examined energy-pumping-based desynchronization in thermosensitive systems. Tagne et al. \cite{Tagne2022} combined light and thermal effects in a photosensitive thermosensitive FHN model, demonstrating bifurcation-induced transitions. Shi et al. \cite{shi2022dynamic} proposed a controllable thermosensitive neuron circuit with parallel thermistors, which revealed diverse dynamic regimes, including periodic, bursting, and chaotic firing. Xu and Ma \cite{xu2022pattern} explored the formation of temperature-dependent patterns in neural networks. Additional contributions include the design of light–temperature adaptive neurons \cite{song2024light} and the investigation of coexisting attractors in Josephson junction circuits driven by temperature and light \cite{ramakrishnan2024coexisting}. These works collectively reflect the growing interest in the multimodal modulation of neuronal excitability.

Despite these advances, the combined influence of temperature and electric fields in a unified framework remains insufficiently explored. Previous models have primarily focused on thermal effects \cite{xu2020dynamics} or electric fields \cite{ma2019model}, overlooking the potential interactions between them in shaping firing dynamics. This gap limits our ability to understand and predict neural behavior under complex physiological conditions.

In this work, we propose a three-variable extension of the FitzHugh–Nagumo model that incorporates both thermosensitivity and the influence of external electric fields. This model enables the simultaneous analysis of intrinsic neuronal dynamics and environmental modulation. Through numerical simulations, we investigate how temperature, cell size, and periodic electric fields shape firing patterns, including transitions between tonic spiking, bursting, and chaotic activity. We use bifurcation diagrams, Lyapunov exponents, and the coefficient of variation (CV) of interspike intervals to characterize the system's behavior under varying parameter regimes. We also examine synchronization in a pair of coupled thermosensitive neurons, showing how coupling through the membrane potential and the electric field modulates coherence. These results offer insights into how multiple environmental factors jointly regulate neuronal excitability and may support future neuromodulation strategies. The remainder of the paper is structured as follows: Section~\ref{sec:model} introduces the model inspired by \cite{ma2019model,hou2021estimate} and analyzes its steady-state dynamics. In section~\ref{sec:results}, we present simulation results on firing behavior under various external conditions, including the combined influence of electric fields and temperature. After that, we explore synchronization in coupled thermosensitive neurons. Section~\ref{sec:discussion} discusses the biological relevance and implications of our findings. Finally, Section~\ref{sec:conclusion} concludes the paper.

\section{Model}

\label{sec:model}
In a neuron,there are numerous charged ions, including calcium, potassium, and sodium,in  movement. Some ions cross the membrane channels, generating transmembrane currents. This movement induces fluctuations in the membrane potential, treating the membrane as a charged surface with a uniform distribution of charges. This ongoing ion flow creates an electric field around the cell, similar to that of a large charged plate. Assuming the membrane has a size $S$ and charge number $q$, the surface charge density $(\sigma = q/S)$ can be calculated, allowing the determination of the electric field intensity near the membrane as follows:
\begin{equation}
	\begin{cases}
		E=\frac{q}{2\epsilon_1 S}=\frac{\sigma}{2\epsilon_1}\\
		\Delta V=rE \approx E\sqrt{S}
	\end{cases}
\end{equation}
where parameter $\epsilon_1$ denotes the dielectric constant that is associated with the intrinsic property of the medium, $r$ is the radius size when the cell is regarded as a ball shape, $\Delta V$ represents the voltage between plates or the membrane potential of the cell, and will be replaced by the variable $V$ in studies. 
As a result, the intensity of the electric field could vary in time during the fluctuation of membrane potential induced by the flow of charged ions across the channels embedded in the membrane. As is well known, biological neuron models should consider the effect of ion channels which decide the propagation of ions and also the membrane potential as well. However, the involvement of the field variable $E$ can well describe the distribution of ions and changes in membrane potential induced by exchange and transport of ions in the cell. Therefore, the electric field can be used as a new variable to estimate the change of ions and the membrane potential of the neuron.\\
Biological neuron models must account for the influence of ion channels, as these govern both ion propagation and membrane potential. By introducing the electric field variable $E$, it becomes possible to characterize the spatial distribution of ions and the dynamic changes of the membrane potential driven by ionic exchange. This approach enriches the description of neuronal dynamics, allowing the inclusion of electric field effects as a critical variable in modeling neural activity\cite{ma2019model}.\\
\quad A thermosensitive neuron model presented by Xu et al.\cite{xu2020dynamics}, is described by:
\begin{equation}
	\begin{cases}
		\frac{dx}{dt}=x(1 - \xi) - \frac{1}{3}x^3 - y + I + A\cos(\omega t) \\
		\frac{dy}{dt} =c[x + a - b\exp(1/T)y]\\ 
	\end{cases}
\end{equation}.

Where $x$ and $y$ represent the membrane potential and the ion current, respectively; $b$ the temperature coefficient, $T$ the temperature, $I$ constant stimulus current, $A$ and $\omega$ the intensity and angular frequency of the external time-varying current, $a$ and $c$ and $\xi$ are constant parameters. So, the effect of the electric field is considered by adding $rE$ to modulate the second variable $y$ which represents the ion current \cite{ma2019model}. Therefore, the improved third-variable thermosensitive FitzHugh-Nagumo model including the effect of the electric field is given as follows.
\begin{equation}
	\begin{cases}
		\frac{dx}{dt}=x(1 - \xi) - \frac{1}{3}x^3 - y + I + A\cos(\omega t) \\
		\frac{dy}{dt} =c[x + a - b\exp(1/T)y]+r E\\
		
		\frac{dE}{dt} =k y \\
	\end{cases}
	\label{eq:syst_without_Eext}
\end{equation}
With $k=\frac{1}{2\epsilon_1 S}$,the excitability of the medium.\\
When neurons are subjected to an external electric field, by the principle of superposition, this field adds to the intrinsic electric field across the membrane. The model is rewritten as:
\begin{equation}
	\begin{cases}
		\frac{dx}{dt}=x(1 - \xi) - \frac{1}{3}x^3 - y + I + A\cos(\omega t) \\
		\frac{dy}{dt} =c[x + a - b\exp(1/T)y]+r E\\
		\frac{dE}{dt} =k y+E_{ext} \\
	\end{cases}
	\label{eq:syst_with_Eext}
\end{equation}  
 Here, we choose $E_{ext}$ as a periodic modulate signal defined by: $E_{ext}=E_m sin(2\pi f t)$ with $E_m$ and $f$ its amplitude and frequency, respectively. However, it is important to note that the stimulus current and external electric fields act through fundamentally different biophysical mechanisms. Specifically, the current term models direct synaptic-like stimulation, influencing the membrane potential via direct current injection, as in current-clamp experiments. In contrast, the external electric field represents an indirect modulation of neuronal activity: it polarizes the membrane and alters the transmembrane potential through capacitive and electrotonic effects. This distinction mirrors realistic physiological conditions in which neurons are simultaneously influenced by synaptic currents and extracellular field fluctuations. Separation of current injection and field modulation has also been adopted in previous modeling studies to reveal their complementary and sometimes interacting effects on neuronal excitability \cite{ma2019model,hou2021estimate}. Including both terms allows us to study their combined effects on excitability and synchronization, as well as their distinct contributions to the overall dynamics of the system.

   To analyze the steady-state dynamics without an external time-dependent drive (\(A=0\), \(E_m=0\)), we set \(\frac{dx}{dt} = \frac{dy}{dt} = \frac{dE}{dt} = 0\). From the third equation, since \(k \neq 0\), it follows that \(y^* = 0\). Substituting this into the second equation gives the equilibrium condition for \(E^*\): 
\begin{equation}
    E^* = -\frac{c}{r} (x^* + a).
\end{equation}
Similarly, the first equation simplifies to  
\begin{equation}
    x^*(1-\xi) - \frac{1}{3} x^{*3} + I = 0,
\end{equation}
which determines \(x^*\).

At equilibrium, \(y\) vanishes due to its direct dependence on \(E\), and the steady-state values of \(x^*\) and \(E^*\) are governed by nonlinear interactions and parameter choices. The cubic term in \(x\) introduces intrinsic self-regulation, while \(E^*\) is modulated by \(x^*\), ensuring a stable fixed-point solution. The positive coupling between \(y\) and \(E\) implies that transient fluctuations in one variable influence the other, but the system ultimately settles in a steady state.

The vanishing current $y^*=0$ is biologically consistent with neurons in a resting or homeostatic state, where net ionic currents are balanced in the absence of external perturbations. Although real neurons exhibit small stochastic fluctuations, our deterministic model captures the essential steady-state behavior without additional noise components.

\section{Numerical simulations }
\label{sec:results}
\quad In this section, the dynamic behaviors of isolated systems of thermosensitive FitzHugh-Nagumo neurons under the electrical field are studied. Numerical simulations were performed using the fourth-order Runge-Kutta method to solve nonlinear differential equations with a time step of 0.01; 
ensuring high precision in capturing the temporal evolution of membrane potential, ion current, and electric field. To analyze the dynamic of the system, we used the coefficient of variation $(CV)$ of the interspikes interval (ISI) to quantify the regularity of neuronal firing. It is computed from the following expression:
\begin{equation}
	CV=\frac{\sqrt{\langle{ISI^2}\rangle-\langle{ISI}\rangle^2 }}{\langle{ISI}\rangle}. 
\end{equation}
Where, $\langle{ISI^2}\rangle$ is the average value of the square of the interval between two spikes and $\langle{ISI}\rangle$ is the average values of the interval of two adjacent spikes. We also used bifurcation analysis to identify transitions between different firing states (quiescent, spiking, bursting, and chaotic) and Lyapunov exponents to detect chaotic behavior by measuring sensitivity to initial conditions. These methods provide a comprehensive understanding of how external stimuli and parameter variations influence neuronal activity, which makes them essential for characterizing neural coding and information processing. 
The fundamental parameters of the thermosensitive FitzHugh–Nagumo model were retained from Xu et al. \cite{xu2020dynamics}, in order to preserve the characteristic excitability and temperature responsiveness observed in previous studies. These parameter choices ensure that the model reproduces biologically consistent transitions between quiescent, spiking, and bursting states under thermal influence.
In contrast, the parameters associated with the external electric field, specifically the cell radius $r$ and the coupling strength $k$, were varied systematically to examine their effects on neuronal dynamics (see Fig.1). The selected range for $r$ reflects typical dimensions of mammalian neurons, while $k$ was adjusted to balance responsiveness and numerical stability. Although the temperature coefficients $b$ and $T$ were initially fixed according to \cite{xu2020dynamics}, we later investigated their dynamic role under electric field modulation (Fig.\ref{7} and Fig.\ref{T_F}), demonstrating their ability to drive transitions between periodic and chaotic regimes. This dual strategy ensured model consistency with prior work while enabling rigorous investigation of the novel electric field effects. 

Thus, the single-neuron dynamic and firing mode are studied with the following selected parameters: $a=0.7$; $c=0.1$; $\xi=0.175$; $b=0.4$; $T=5$,$I=0.5$ with the fixed initial conditions values for the variables ($x_0,y_0,E_0$) =($0.1,0.3,0.003$).
\subsection{Single neuron dynamics in the absence of the external electric field}
 First, the dynamic behavior of the isolated thermosensitive FitzHugh-Nagumo neuron is studied in the absence of the external electric field $E_{ext}$ (equations~(\ref{eq:syst_without_Eext})).

\begin{figure}[h!]
	\centering
	\includegraphics[width=8.5cm,height=6cm]{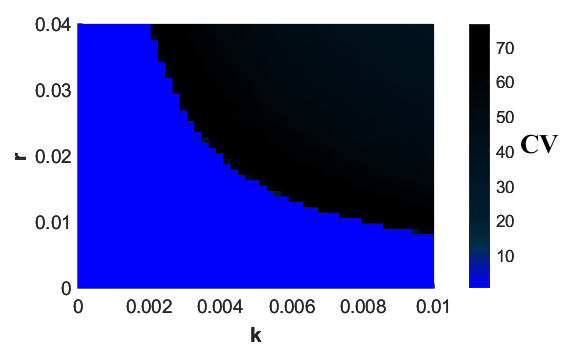}
	\caption{Coefficient of variation of the interspikes interval with cell radius $r$ and excitability parameter $k$ for $w=0.005$ . Periodic bursting occurs in the blue region. The black region represents non-spiking or subthreshold activity. The others parameters are  $a=0.7$; $c=0.1$; $\xi=0.175$; $b=0.4$; $T=5$,$I=0.5$ } \label{1} 
\end{figure}

\begin{figure}[h!]
	\centering
	\includegraphics[width=8cm,height=3cm]{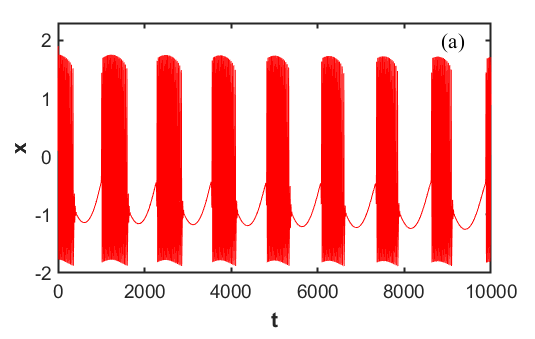} 
	\includegraphics[width=8.1cm,height=3cm]{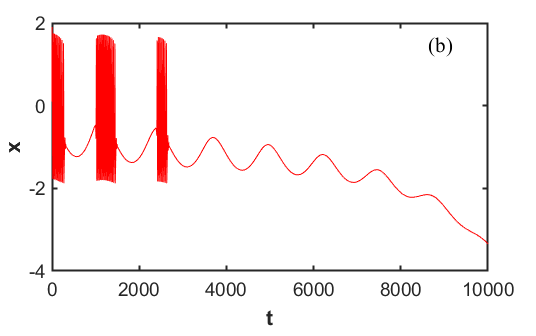}  
	\includegraphics[width=8cm,height=3cm]{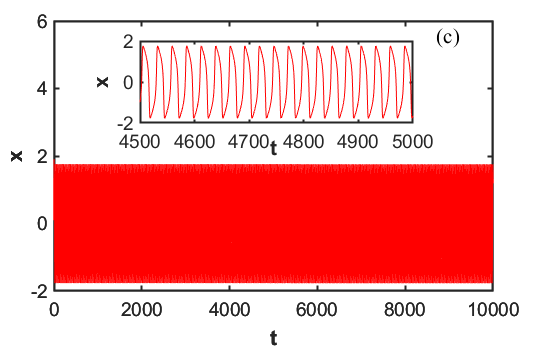} 
	\includegraphics[width=8cm,height=3cm]{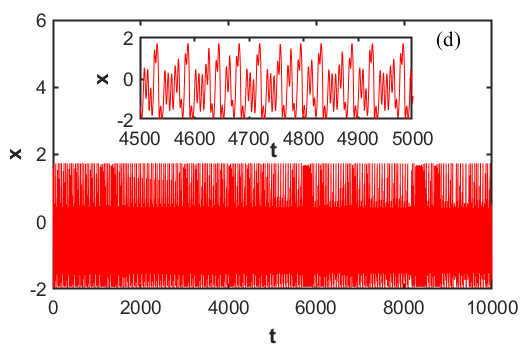}
	\caption{Time series of the variable x for $w=0.005$ a)$r=0.0001$; b)$r=0.06$, and for $r=0.0001$ c)$w=0$; d)$w=1.004$ .The others parameters are $a= 0.7$, $c=0.1$, $\xi=0.175$, $T=5$, $A=0.9$ ,$k=0.001$, $b=0.4$, $I=0.5$ \label{2}}  
\end{figure}
\quad To assess the impact of the scaling of the cell radius $r$ and the excitability parameter $k$, a contour plot of the coefficient of variation $(CV)$ of the interspikes interval is shown in Fig.\ref{1}, with fixed values of the amplitude and angular frequency of the periodic external current stimulus($A=0.9$, $\omega=0.005$). This plot reveals that, in the absence of an external electric field, the $CV $ exhibits clear continuity, particularly when the angular frequency is small. In particular, $CV$ is minimized when both the cell radius $r$ and the excitability parameter $k$ are small. For larger values of $r$ and $k$, the neuron displays transient bursting followed by suppression of firing and eventual divergence. These regions produce artificially high values of the coefficient of variation ($CV>10$)  and are excluded from further dynamical analysis because they are outside the physiologically relevant domain.
 To further elucidate this sensitivity of neuron activity to the cell radius $r$, Fig.\ref{2} presents time series of the membrane potential illustrating the mode dependence of electrical activities on the cell radius. It is evident that neuronal activity is significantly affected by the radius of the neuron cell, especially with an appropriate excitability parameter value. For subsequent analysis, we set the excitability parameter $k=0.001$. Analyzing Fig.\ref{2}.c and Figure \ref{2}.d, we observe that the neuron exhibits, respectively, periodic and chaotic spiking activities depending on the value of the angular frequency of the external stimulus source. 
\begin{figure}[h!]
	\centering  
	\includegraphics[width=8.5cm,height=4.5cm]{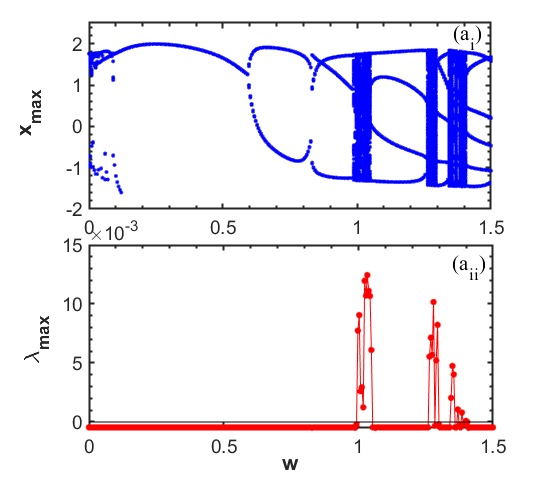}\\
	\includegraphics[width=8.5cm,height=4cm]{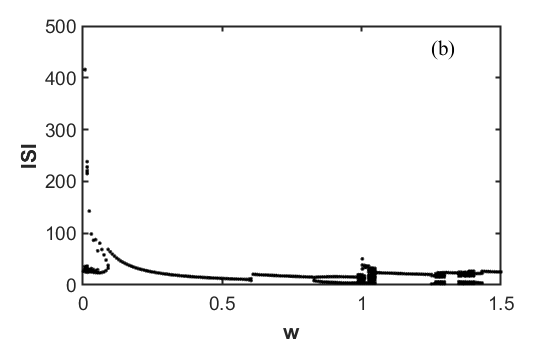} 
	\caption{a)Bifurcation diagram and largest Lyapunov exponent, b) Interspikes interval $ISI$ concerning the angular frequency of the external stimulus  $\omega$ for $A=0.9$ and the radius cell $r=0.0001$.The others parameters are $a= 0.7$, $c=0.1$, $\xi=0.175$, $I=0.5$, $b=0.4$, $T=5$, $k=0.001$\label{3}} 
\end{figure}      
Figure\ref{3}.a presents the bifurcation diagram of neuronal activity as a function of external frequency $\omega$, along with the corresponding largest Lyapunov exponent $\lambda_{\max}$. It is depicted that by changing the angular frequency $\omega$, neurons present  different regime(periodic, multi-periodic and chaotic). In fact, for lower values of $\omega$, the neuronal response remains periodic, as indicated by the distinct branches in the bifurcation graph and the non positive Lyapunov exponents. However, as $\omega$ increases, complex bifurcations emerge, leading to chaotic dynamics characterized by positive values of $\lambda_{\max}$. Fig.\ref{3}.b shows the distribution of the interspikes interval $(ISI)$ as a function of $\omega$. For small values of $\omega$, the values of $ISI$ are large and highly variable, indicating slow periodic or bursting activity. As $\omega$ increases, $ISI$ decreases, reflecting an increase in firing frequency. Around the transition points observed in the bifurcation diagram, fluctuations in $ISI$ are evident, suggesting a transition from periodic to chaotic or irregular bursting behavior.
\begin{figure}[h!]
	\centering
    \includegraphics[width=8cm,height=3cm]{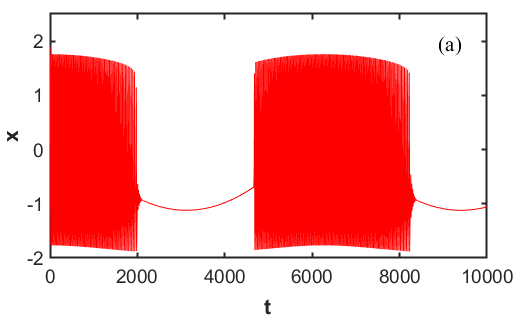}\\
    \includegraphics[width=7.8cm,height=3cm]{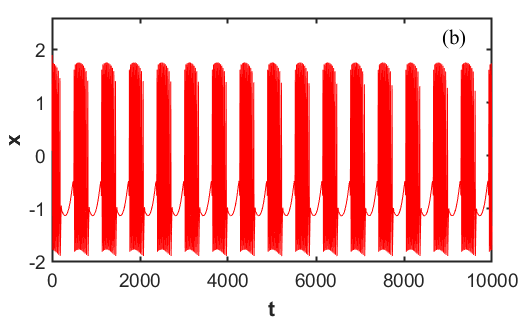} 
    \includegraphics[width=7.8cm,height=3cm]{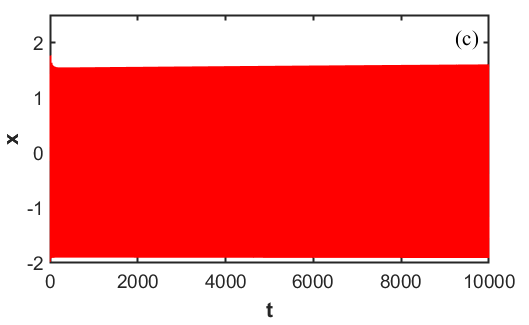} 
    \includegraphics[width=8cm,height=3cm]{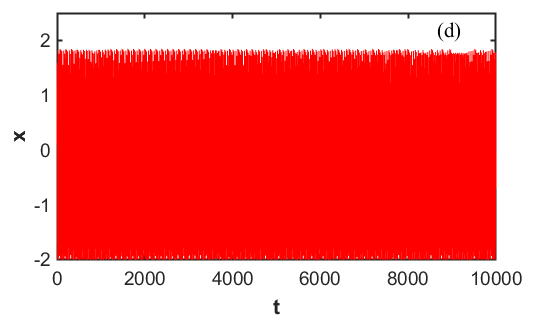} 
    
	\caption{Time series of the membrane potential under periodic external stimulus $Acos(wt)$ at different frequency : a)$w=0.001$, b)$w=0.01$, c)$w=0.6$, d)$w=1.27$. With  the others parameters $A=0.9$,$k=0.001$,$r=0.0001$,$a=0.7$,$c=0.1$,$\xi=0.175$,$I=0.5$,$b=0.4$,$T=5$ \label{4}}
\end{figure} 

\quad  The evolution of neuronal firing over time at different angular frequencies is presented in Fig.\ref{4}. The result shows that the firing mode is affected by the angular frequency of the external stimulus source with fixed amplitude. Then, the appropriate external stimulus can effectively change the firing dynamic or the excitability of the neuron.
 \begin{figure}[h!]
	\includegraphics[width=8cm,height=6cm]{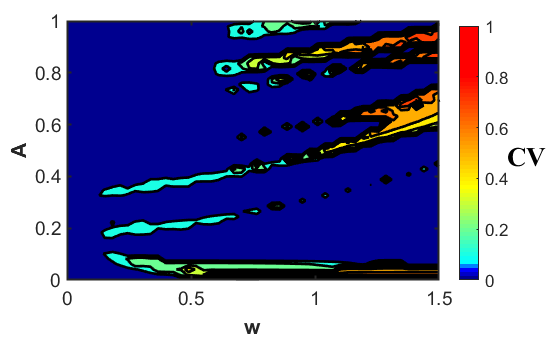} 
	\caption{Coefficient of variation of the interspikes interval with the external stimulus intensity $A$ and the  angular frequency  $\omega$ with $k=0.001$, $r=0.0001$ , $a= 0.7$,$c=0.1$,$\xi=0.175$,$I=0.5$,$b=0.4$,$T=5$ \label{5} } 
\end{figure}
\begin{figure}
    \includegraphics[width=8cm,height=3cm]{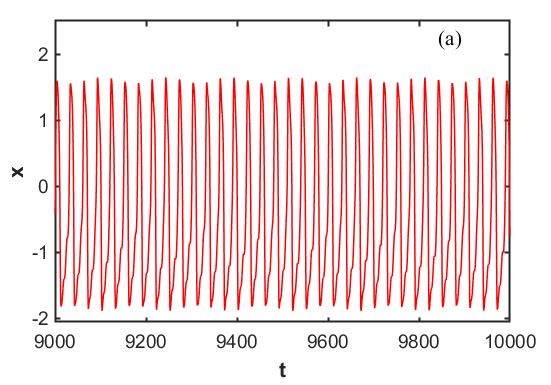}
    \includegraphics[width=8cm,height=3cm]{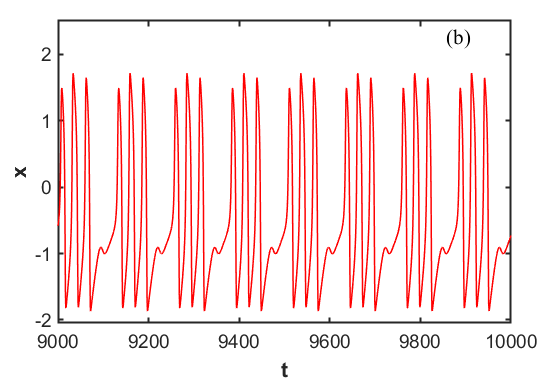}
	\includegraphics[width=8cm,height=3cm]{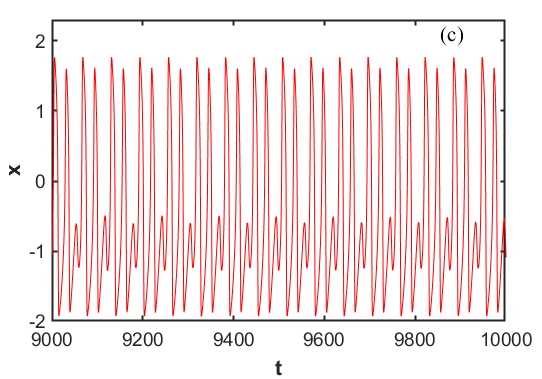}\\
    \includegraphics[width=8cm,height=3cm]{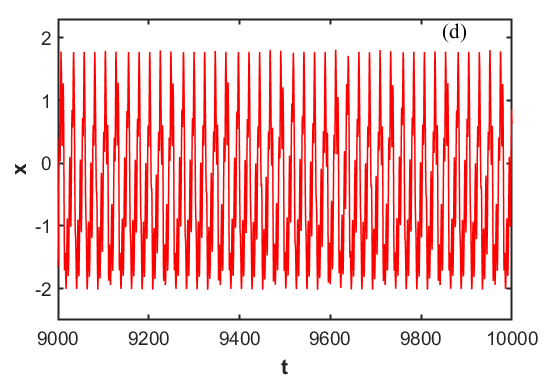}
	\caption{Time series of the membrane potential under alternative external stimulus current  at different intensity and a angular frequency: a)$A=0.1$,$w=1.004$, b)$A=0.4$, $w=0.05$, c)$A=0.3$,$w=0.2$, d)$A=9$,$w=1.35$.\label{fig6}The others parameters are $a= 0.7$,$c=0.1$,$\xi=0.175$,$I=0.5$,$b=0.4$,$T=5$ $r=0.0001$}
\end{figure}

\quad Fig.\ref{5} shows the relationship between the amplitude and the angular frequency of the external stimulus current for the radius cells of$r=0.0001$. It shows that the coefficient of variation takes values less than 1 but there are various mutation values. It should be noted that $CV$ is the smallest when the angular frequency or the amplitude is small, indicating regularity in the spike intervals.
Figure~\ref{5} displays a contour plot of the coefficient of variation $(CV)$ of the interspikes interval as a function of the amplitude $A$ and frequency $\omega $ of the external periodic stimulus current. The figure reveals complex firing dynamics across the $(\omega, A)$ parameter space. At low frequency $\omega$, the neuron displays more regular firing over a broad amplitude range, as evidenced by the presence of blue regions (low $CV$). However, as $\omega$ increases, different regions of high $CV$ emerge, indicating transitions to irregular or chaotic spiking behavior. These irregular zones appear interspersed with pockets of low $CV$, suggesting the presence of frequency-dependent resonance-like effects where the electric field entrains the firing into regular patterns. This structure illustrates the neuron's nonlinear response to sinusoidal current input, where small changes in the stimulus parameters can lead to qualitative changes in spiking behavior.

For further illustration, Fig.\ref{fig6} presents neuronal activity over time for the external voltage amplitude ($A$) and frequency ($\omega$) chosen. The corresponding coefficients of variation (CV) values are 0.01, 0.02, 0.1, and 0.8, respectively.
In Fig.\ref{fig6}.a, the low variability ($CV \approx 0.01$) suggests tonic spiking , characterized by stable and rhythmic firing. In Fig.\ref{fig6}.(b,c), $CV \approx 0.02$, $CV \approx 0.1$, respectively, suggest a transition to bursting with slight modulations in spike timing, where variations in burst duration and interburst intervals emerge. In contrast, Fig.\ref{fig6}.d shows highly irregular firing ($CV \approx 0.8$), indicating chaotic spiking, where loss of periodicity leads to erratic neuronal activity. 
\begin{figure}
	\centering
	\includegraphics[width=8.5cm,height=6cm]{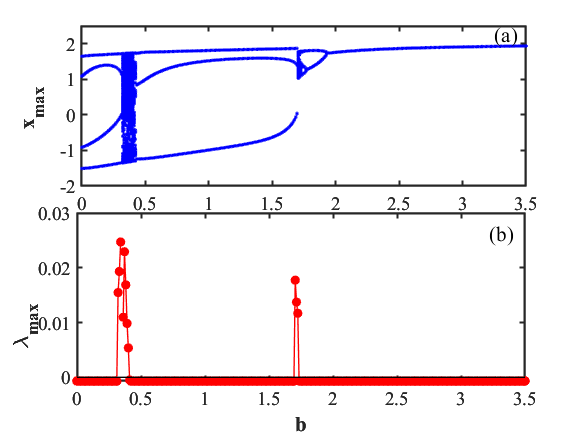} 
	\caption{Bifurcation and Largest Lyapunov exponent varying $b$ for $r$ fixed at $0.0001$
	}
	\label{7}
\end{figure}
\quad In the thermosensitive model proposed in \cite{xu2020dynamics}, the coefficient of temperature $b$ plays a crucial role in the dynamic of individual neurons. To investigate its effect considering the intrinsic electrical field, the bifurcation diagram and the maximal Lyapunov exponent are represented in Fig.\ref{7}. It is shown that, for different values of this coefficient of temperature, the system present different dynamic regime like: periodic, multiperiodic and chaotic dynamic.\\
\begin{figure}[h!]
	\centering
	\includegraphics[width=8cm,height=6cm]{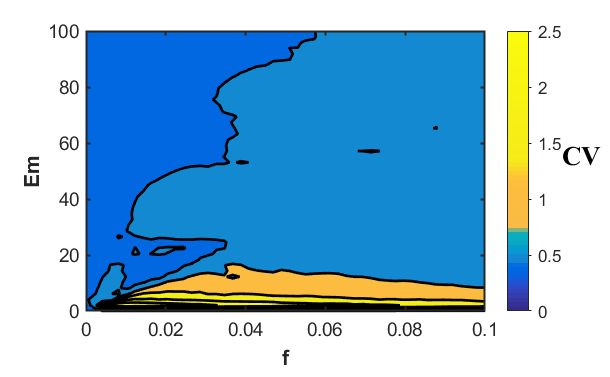}
	\caption{Coefficient of variation of the interspikes interval with the external stimulus intensity $Em$ and the  frequency $f$ for a) $r=0.0001$, $a=0.7$,$c=0.1$,$\xi=0.175$,$I=0.5$,$b=0.4$,$T=5$,$w=1.004$,$A=0.9$ \label{8} } 
\end{figure}
\begin{figure}
	\centering   
    \includegraphics[width=8cm,height=3.5cm]{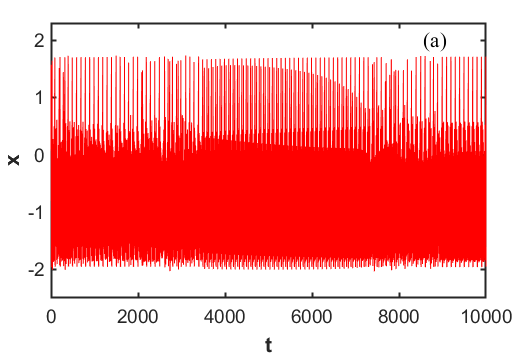}
     \includegraphics[width=8cm,height=3.5cm]{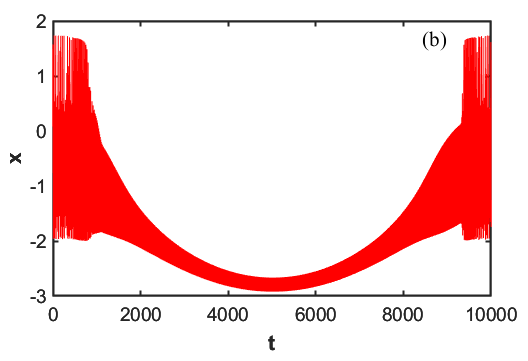}
      \includegraphics[width=8cm,height=3.5cm]{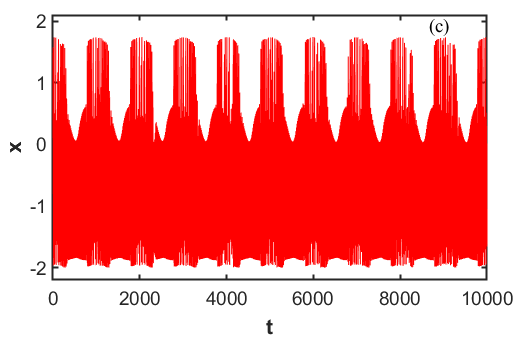}
       \includegraphics[width=8cm,height=3.5cm]{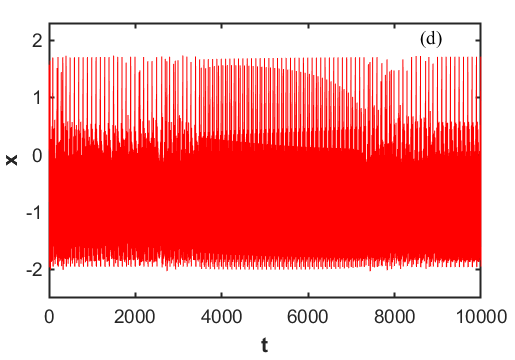}
	
	\caption{Time series of the membrane potential under alternative external stimulus  at different frequency: a)$f=0$,$Em=1.5$, b)$f=0.0001$,$Em=1.5$, c)$f=0.001$, $r=0.0001$, $Em=1.5$, d)$f=0.1$,$Em=40$.The others parameters are $a=0.7$, $c=0.1$, $\xi=0.175$, $I=0.5$, $b=0.4$, $T=5$,  $\omega=1.004$, $A=0.9$, $r=0.0001$ \label{9}} 
\end{figure}

\subsection{Effect of the external electric field}
When a neuron is exposed to an external field, its electrical activity is altered. For a sinusoidal external electric field, the amplitude and frequency are two critical parameters (see Equation~\ref{eq:syst_with_Eext}). The coefficient of variance $(CV)$ is used to analyze the effect of the amplitude $E_m$ and the frequency $f$ of the external electric field on the neuronal firing pattern with a relative high external current frequency $(w=1.004)$. In Fig.\ref{8}, the contour plot reveals that neuronal firing is highly regular (low CV values, blue region) over a wide range of amplitudes and frequencies of the external field, particularly at higher amplitudes ($E_m > 20$) and low frequencies. A transition zone appears around $E_m \approx 20 $ and $f \approx 0.01$, where the values of $CV$ gradually increase, indicating more variability in the firing pattern. The highest variability (yellow region) is observed for low frequencies ($f \leq 0.05 $) and moderate amplitudes ($E_m \leq 20$), where the external field has the strongest effect on neuronal dynamics, resulting in irregular firing behavior. In general, the system displays regular firing at higher amplitudes, while moderate frequencies and low amplitudes introduce complex, irregular dynamics.

\quad Indeed, the external electric field modulates the amplitude of the membrane potential and also the neuronal firing mode with an appropriate cell size value. Several cases illustrating the evolution of the neuronal membrane potential over time are depicted in Fig.\ref{9}. In Fig.\ref{9}.a ($f = 0$), no external electric field is applied; the neuron exhibits irregular, chaotic-like activity due to the intrinsic dynamics of the system at $\omega = 1.004$. In Fig.\ref{9}.b ($f = 0.0001$, $E_m = 1.5$), the introduction of a very low-frequency electric field begins to modulate the neuron, creating subtle timing changes in spike generation. This represents the onset of field-driven modulation, although intrinsic irregularity still dominates. In Fig.\ref{9}.c ($f = 0.001$), the field effect becomes more evident. The neuron transitions toward partially entrained behavior: the spike train shows periodic modulation, and the variability of interspike intervals decreases. This reflects the system’s entry into a more coherent dynamical regime influenced by the external field. Finally, in Fig.\ref{9}.d ($f = 0.1$, $E_m = 40$), the neuron demonstrates regular, high-frequency spiking, characteristic of complete entrainment. Here, the electric field dominates the system, overriding intrinsic chaotic dynamics and inducing a tonic spiking pattern with nearly constant interspike intervals.\\
\begin{figure}[h!]
	\centering       
	\includegraphics[width=8.5cm,height=6cm]{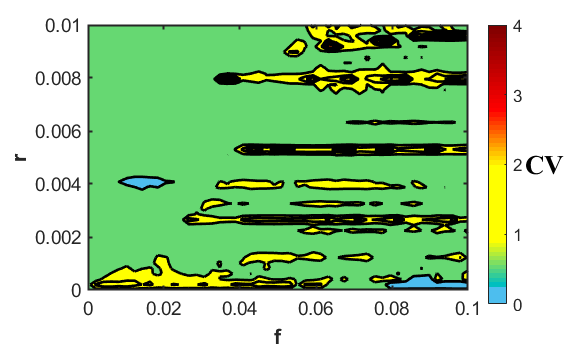} 
	\caption{coefficient of variation with $r$ and $f$ with  $\omega =1.004$, $E_m=1.5$. The others parameters are $a=0.7$, $c=0.1$, $\xi=0.175$, $I=0.5$, $b=0.4$, $T=5$, $A=0.9$  \label{10}} 
\end{figure}
\begin{figure}[h!]
	\centering    
    \includegraphics[width=8cm,height=3cm]{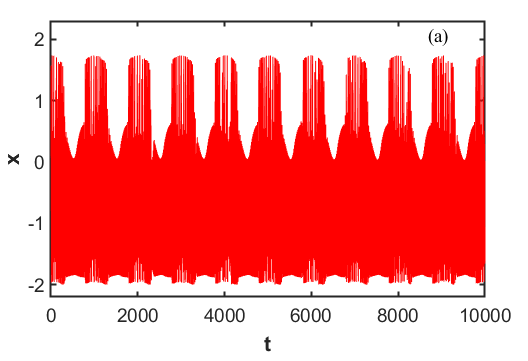}
    \includegraphics[width=8cm,height=3cm]{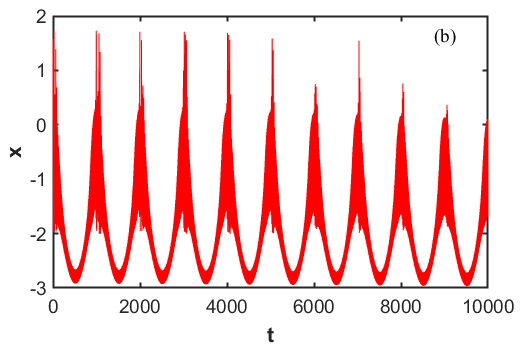}
    \includegraphics[width=8cm,height=3cm]{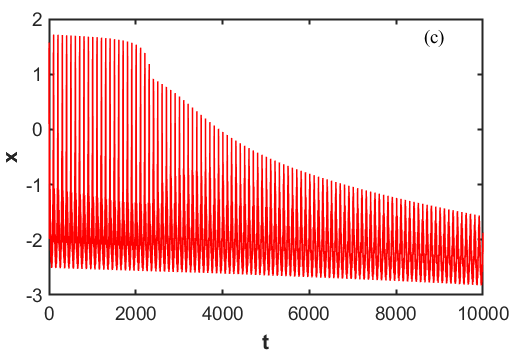}
    \includegraphics[width=8cm,height=3cm]{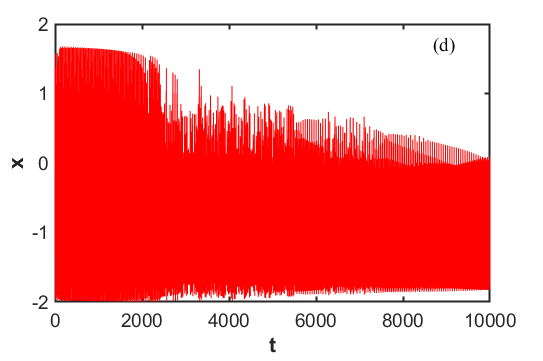}
	
	\caption{Time series of the membrane potential under alternative external stimulus  at different frequency for $Em=1.5$ and  $\omega =1.004$: a)$f=0.002$, $r=0.0002$, b)$f=0.001$, $r=0.001$, c)$f=0.015$, $r=0.007$, d)$f=0.1$, $r=0.007$.The others parameters are $a=0.7$, $c=0.1$, $\xi=0.175$, $I=0.5$, $b=0.4$, $T=5$,  $\omega =1.004$, $A=0.9$  \label{11}} 
\end{figure}      
\quad \quad  Now we investigate the impact of an external field frequency on the spiking-bursting activity of neurons, focusing on the chaotic regime ( $\omega  = 1.004$) and weak electric field intensity ($Em=1.5$) . In Fig.\ref{10}, a contour plot of the coefficient of variation $(CV)$ of interspikes intervals is presented against the cell radius and external electric field frequency. The blue and green light areas indicate regions where the coefficient of variation is small (less than 1 and closer to 0), meaning that the neuron firing is more regular. This tends to happen at lower field frequencies $f<0.002$ and specific radii, especially around $r\approx0.004$ and slightly below $r=0.002$. The yellow to red areas indicate regions where the neuronal firing is more irregular or variable, with the CV increasing. These regions are observed in a range of frequencies and radii, particularly noticeable for $r>0.004$  and in different frequency bands ($f\approx0.03$ to $f\approx 0.08$). The light green areas in the plot mark transitions between more regular firing (blue) and highly irregular or bursty firing (yellow/red). In fact, around $f \approx 0.04$ and $ r \approx 0.004 $: This is one of the larger light green regions, indicating that at this combination of frequency and cell radius, the firing pattern has a moderate level of variability. Spanning from  $f \approx 0.01$  to  $f \approx 0.05 $ for  $r \approx 0.002$, the neuron activity transitions from highly regular (blue) to more irregular (yellow) as frequency increases, but stays in a light green zone for a certain range, suggesting some complex dynamics.
The plot shows elongated regions where the $CV$ is higher, suggesting that certain combinations of external field frequencies and cell radii lead to more bursty or irregular neuron firing patterns.  It is more illustrated in Fig.\ref{11}, where the temporal evolution of the membrane potential variable is presented for different values of $f$ and $r$. In particular, Fig.\ref{11}.c and Fig.\ref{11}.d illustrate a transition to small-amplitude oscillations under relatively high values of cell radius $r$ and stimulation frequency $f$. Under these conditions, the membrane of the neuron becomes increasingly influenced by external periodic input and its capacitive effects. A larger radius $r$ enhances the surface area exposed to the field or current, while a high-frequency stimulus can desynchronize or suppress the intrinsic firing rhythm. As a result, the system may enter a regime where oscillations are damped, and the trajectory converges to a stable, low-amplitude periodic orbit or a steady state; consistent with amplitude death phenomena observed in nonlinear excitable systems.
This transition reflects how the combined influence of geometry and fast stimulation can suppress neural excitability. 
\\

\begin{figure}[h!]
	\centering       
	\includegraphics[width=8.5cm,height=6cm]{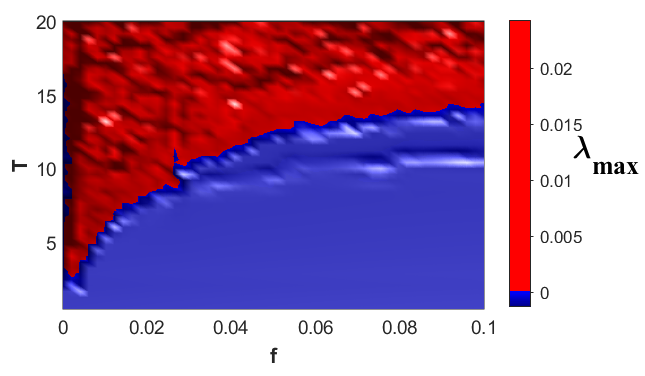} 
	\caption{Lyapunov Exponent against $T$ and $f$ with $b=0.47$ , $E_m=1.5$ presenting chaotic dynamic region(red) and regular dynamic region (blue). The others parameters are $a=0.7$, $c=0.1$, $\xi=0.175$, $I=0.5$,  $\omega =1.004$, $A=0.9$, $r=0.0001$  \label{T_F}} 
\end{figure}

To explore how the interaction between temperature ($T$) and external electric field frequency ($f$) influences neuronal dynamics, we analyzed the maximum Lyapunov exponent ($\lambda_{\max}$) in different parameter regimes (Fig.~\ref{T_F}). Fixing $b = 0.47$, we observed that at low temperatures ($T < 4$), neuronal activity remains regular ($\lambda_{\max} \leq 0$, blue regions). However, as $T$ increases, the system undergoes a transition to chaos ($\lambda_{\max} > 0$, red regions), indicating increased neuronal excitability. Furthermore, the frequency of the external field modulates these transitions. At low $f$, the chaotic and regular regimes co-exist, while at higher $f$, chaotic activity dominates with localized stability windows. This interaction between $T$ and $f$ suggests that external fields can modulate temperature-driven chaotic transitions, providing novel insights into neuronal excitability control. 
\subsection{Synchronization of two coupled neurons}
In this section, we explore the synchronization of two coupled thermosensitive FitzHugh-Nagumo neurons under an electric field. The model is: 
\begin{equation}
	\begin{cases}
		\frac{dx1}{dt}=x1(1 - \xi) - \frac{1}{3}x1^3 - y1 + I + A\cos(\omega t)+g1(x2-x1) \\
		\frac{dy1}{dt} =c[x1 + a - b\exp(1/T)y1]+r E1\\
		
		\frac{dE1}{dt} =k y1+E_{ext}+ g2(E2-E1) \\
		\frac{dx2}{dt}=x2(1 - \xi) - \frac{1}{3}x2^3 - y2 + I + A\cos(\omega t)+g1(x1-x2) \\
		\frac{dy2}{dt} =c[x2 + a - b\exp(1/T)y2]+r E2\\
		
		\frac{dE2}{dt} =k y1+E_{ext}+ g2(E1-E2)\\ 
		
	\end{cases}
\end{equation}  
Here, the coupling strengths between neurons through the membrane potential and the electric field are $g1$ and $g2$, respectively. To quantify synchronization between two neurons,  the average synchronization error is introduced $Er = \frac{1}{T}\int_{t}^{t+T} e(t)dt$, where {\small{$e(t)=\sqrt{(x2 - x1)^2 + (y2 - y1)^2 + (E2 - E1)^2 }$}} the instantaneous error and \( T \) is the total simulation time. 
\begin{figure}[h!]
	\centering       
	\includegraphics[width=8.5cm,height=6cm]{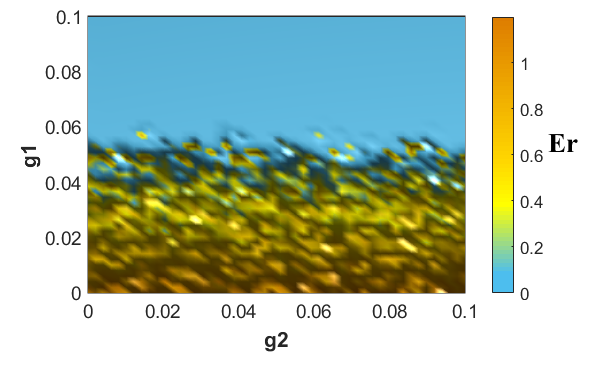} 
	\caption{ Average synchronization error $Er$ against $g1$ and $g2$ with  $\omega =1.004$,$E_m=1.5$,$f=0.01$
		$a=0.7$, $c=0.1$, $\xi=0.175$, $I=0.5$,$b=0.4$, $T=5$, $A=0.9$. the initial conditions are set as follows:($x1$,$y1$,$E1$,$x2$,$y2$,$E2$)=(0.1,0,0,0,0,0)   \label{12}} 
\end{figure}

\quad In Fig.\ref{12}, the average synchronization error $Er$ between two coupled neurons is shown as a function of the coupling strength through the membrane potential variable $ g_1$ and the coupling strength through the electric field variable $ g_2$. The color scale represents the synchronization error, with darker green areas indicating high errors (poor synchronization) and yellow regions corresponding to low errors (better synchronization).

The plot reveals that synchronization is highly sensitive to variations in both $g1$  and $ g_2$. At low values of $g1$  and $ g_2$ , the synchronization error is maximal (dark yellow), indicating week-long synchronization between neurons when both coupling strengths are weak. As  $g1$   increases, particularly in the upper regions where $g_1 > 0.05$, the error decreases (blue), suggesting that synchronization becomes stronger in coupling strengths, especially through the membrane potential variable. Interestingly, there are regions in the midrange of $g1$ and $ g_2$ where asynchronization is enhanced, as seen in the small, scattered yellow areas ( $g1 \approx 0.04$  to  $g1 \approx 0.058$ . These regions likely correspond to parameter combinations where the interaction between the membrane potential and electric field coupling creates asynchronization effects. However, in general, better synchronization is observed in high regions of coupling strengths $g1$. 
\begin{figure}[h!]
	\centering     
	\includegraphics[width=8cm,height=3.5cm]{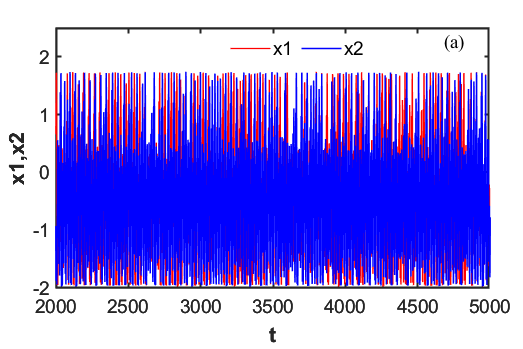}\\
	\includegraphics[width=8cm,height=3.5cm]{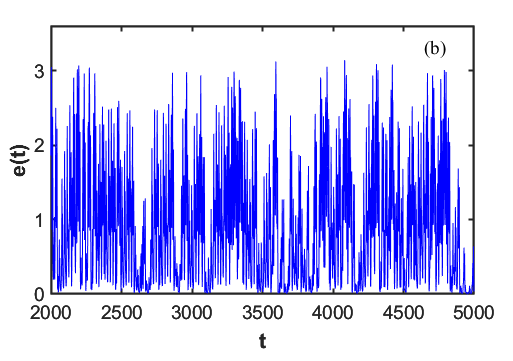}    
	\includegraphics[width=8cm,height=3.5cm]{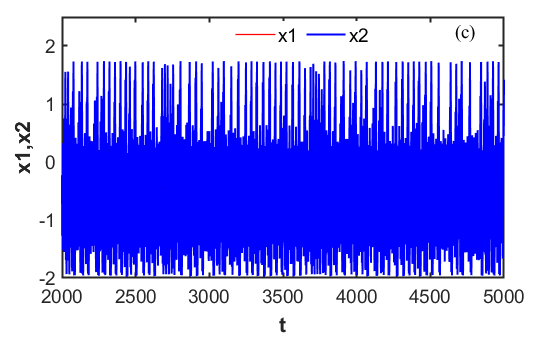}      \includegraphics[width=8cm,height=3.5cm]{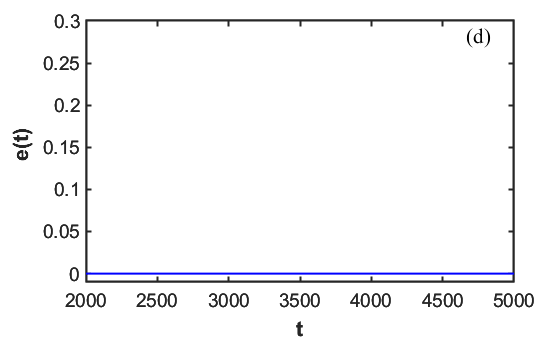}
	\caption{Time series of the membrane potential and the corresponding synchronization error: (a,b) asynchronous state for $g1=0.01$ and $g2=0.02$;(c,d)synchronous state for $g1=0.06$ and $g2=0.04$.The others parameters are $a=0.7$, $c=0.1$, $\xi=0.175$, $I=0.5$,$b=0.4$, $T=5$,  $\omega =1.004$, $A=0.9$.\label{13}} 
\end{figure}
For more illustrations,the firing activities and the corresponding synchronization error are shown in Fig.\ref{13}. Here we present two cases, the first case where the two neurons behave asynchronously over time for $g1=0.01$ and $g2=0.02$, as is confirmed by the synchronization error(see Fig.\ref{13}(a,b)). The second case, where coupled neurons show a firing synchronous state for $g1=0.06$ and $g2=0.04$ (see Fig.\ref{13}(c,d)). 

\section{Discussion}
\label{sec:discussion}
We investigated the dynamics of a thermosensitive FitzHugh-Nagumo neuron under combined temperature and electric field modulation. Our results reveal an intricate interplay between thermal effects, excitability, and external field parameters in controlling neuronal firing patterns.

We found that the analysis of the coefficient of variation $(CV)$ shows that neuronal activity strongly depends on cell radius $r$ and excitability parameters. Small values of these parameters maintain highly regular spiking patterns with minimal variability (Fig.~\ref{1}), demonstrating how neuronal size and excitability govern spike generation robustness in biological systems.

Our bifurcation analysis and Lyapunov exponent calculations demonstrate that external stimuli drive the neuron through various dynamic regimes, periodic, multiperiodic, and chaotic spiking, depending on angular frequency $\omega$ (Fig.~\ref{3}). The analysis of $CV$ (Fig.~\ref{5}) confirms this behavior, showing that a high value of $\omega$ and stimulus intensity $A$ increase firing irregularity, with a nonmonotonic dependence on $A$ at a fixed high $\omega$.

We examined electric field effects ($E_m$, $f$) separately and discovered that they modulate synchronization differently than direct stimuli. High $E_m$ with low $f$ maintains regular firing, while moderate values induce complex patterns (Fig.~\ref{8}). $CV$ analysis (Fig.~\ref{10}) reveals radius-dependent responses, with regular spiking dominating at $r \approx 0.004$ and lower frequencies.

Our simulation (Fig.~\ref{7}) shows that temperature variations through coefficient $b$ significantly alter neuronal dynamics. Increasing $b$ leads to transitions between regular and chaotic dynamics, supporting the hypothesis that temperature fluctuations such as those occurring during fever or metabolic changes\cite{saravia2006hippocampal,yellen2018fueling,dube2005interleukin}; affect neuronal excitability and synchronization.\cite{kim2012high,volgushev2000synaptic}

Through Lyapunov exponent analysis (Fig.~\ref{T_F}), we quantified how temperature $T$ and field frequency $f$ jointly modulate chaotic dynamics. We observed three distinct regimes:
\begin{itemize}
\item Low temperatures ($T < 4$): Regular dynamics dominate
\item Intermediate temperatures: Mixed chaotic/regular states emerge
\item High temperatures: Fully chaotic behavior develops
\end{itemize}
The bifurcation diagram (Fig.~\ref{7}) confirms these transitions, showing how increasing $b$ induces multiple stability changes.

Unlike previous studies that examined thermosensitivity and fields separately~\cite{ma2019model,xu2020dynamics}, our unified model uncovers a broader range of neuronal behavior. The control parameters ($r, k$) we introduced refine firing transition characterization,and our analyze reveals that external fields reshape rather than simply induce or suppress chaotic activity.

Our findings demonstrate that intrinsic properties (size, excitability, temperature) and external modulations (stimuli, fields) interact complexly to shape neuronal responses. This systematic parameter mapping advances our understanding of neuron-field interactions for neuromodulation applications.

We acknowledge three key limitations in this study. First, our deterministic model neglects neurobiological noise sources such as thermal fluctuations and stochastic channel gating. Second, the equilibrium condition $y^*=0$  oversimplifies real neuronal activity. Third, while the FitzHugh-Nagumo model provides valuable insights, it lacks biological details like voltage-gated channel kinetics. Future work should incorporate stochastic elements to bridge this gap.
\section{Conclusions}
\label{sec:conclusion}
In this work, we proposed and analyzed a three-variable thermosensitive FitzHugh–Nagumo neuron model that integrates both intrinsic neuronal dynamics and the effects of external electric fields. By combining bifurcation analysis, Lyapunov exponents, and spike-based metrics such as the coefficient of variation $(CV)$ of interspikes intervals, we investigated how thermal conditions, cell radius, and external stimulation jointly shape neuronal firing patterns. Our results revealed a rich repertoire of dynamical regimes, including tonic spiking, bursting, chaotic activity, and quiescence, along with transitions between these states.

Importantly, we demonstrated that the cell radius plays a critical role in modulating the neuron’s exposure to external fields, which in turn influences excitability and firing regularity. This effect is linked to piezoelectric phenomena, where changes in cell size affect membrane capacitance and the time-varying currents induced by external fields. Furthermore, we examined the synchronization behavior in a pair of coupled thermosensitive neurons, showing how electric field-mediated interactions and coupling strength modulate coherence between units.

Our findings are consistent with previous studies~\cite{xu2020dynamics,shi2022dynamic,Tagne2022}, which investigated the dynamics of thermosensitive neurons under various conditions; and advance the field by explicitly incorporating external electric fields and analyzing synchronization in coupled neurons dimensions, not fully addressed in earlier work.

The proposed framework improves our understanding of how environmental factors regulate neuronal excitability and offers insights that can inform neuromodulation strategies that involve thermal or electrical stimulation.
Looking ahead, this approach can be extended to larger networks of thermosensitive neurons with complex topologies to study phenomena such as partial synchronization and chimera states. Future research could also incorporate additional factors, such as magnetic fields or synaptic interactions, to deepen our understanding of the complex dynamics shaping neural systems.
\\
\\
\noindent \textbf{Acknowledgment} \hspace{3mm} ACR thanks Sao Paulo Research Foundation\\ (FAPESP) . FFF and ACR thank Brazilian National Council for Scientific and Technological Development (CNPq). ELFN thanks Brazilian Federal Agency for Support and Evaluation of Graduate Education (CAPES).
\\
\\

\noindent \textbf{ Funding declaration } \hspace{3mm}

This work was supported by the Sao Paulo Research Foundation (FAPESP, grant 2013/07699-0), the Brazilian National Council for Scientific and Technological Development (CNPq, grants 303359/2022-6 and 316664/2021-9), and the Brazilian Federal Agency for Support and Evaluation of Graduate Education (CAPES, grants 001).
\\
\\
\noindent \textbf{ Author Contribution }
E.L.F.N., F.F.F., and A.C.R. contributed to the conceptualization, methodology, and formal analysis.
E.L.F.N. conducted the investigation, performed simulations and analysis of neuronal dynamics, and prepared all figures and visualizations.
E.L.F.N. and F.F.F. wrote the original draft.
E.L.F.N., F.F.F., and A.C.R. reviewed and edited the manuscript.
F.F.F. and A.C.R acquired funding.
F.F.F. supervised the project and managed administration.
All authors reviewed the manuscript.
\\

\noindent \textbf{ Data availability } No datasets were generated or analyzed during the current study.
\\
\\
\noindent \textbf{ Conflict of interest }The authors declare that they have no conflict of interest. 


\bibliographystyle{unsrt}
\bibliography{references}

\end{document}